\documentclass[amsmath,amssymb,aps,prb,onecolumn,superscriptaddress]{revtex4-2}
\usepackage[utf8]{inputenc}
\usepackage[normalem]{ulem}
\usepackage[colorlinks]{hyperref}
\hypersetup{colorlinks=true,
linkcolor=blue,
filecolor=blue,
urlcolor=blue,
citecolor=blue}
\usepackage{epsfig,subfigure,float,xcolor}
\newcommand\beq{\begin{equation}}
\newcommand\eeq{\end{equation}}
\newcommand\bes{\begin{subequations}}
\newcommand\ees{\end{subequations}}
\newcommand\bea{\begin{eqnarray}}
\newcommand\eea{\end{eqnarray}}
\newcommand\non{\nonumber}

\newcommand\ig{\includegraphics}

\newcommand\al{\alpha}

\newcommand\ga{\gamma}

\newcommand\ep{\epsilon}

\newcommand\la{\langle}
\newcommand\ra{\rangle}

\newcommand\Om{\Omega}

\newcommand\dg{\dagger}

\newcommand\da{\downarrow}
\newcommand\sfig{\subfigure}

\newcommand\vk{{\bf k}}

\newcommand{\mc}{\mathcal}
\newcommand{\mfr}{\mathfrak}
\newcommand{\RNum}[1]{\uppercase\expandafter{\romannumeral #1\relax}}

\begin{document}
\title{Floquet topological phases on a honeycomb lattice using elliptically polarized light}
\author{Ranjani Seshadri}
\email{ranjanis@post.bgu.ac.il}
\affiliation{Department of Physics, Ben-Gurion University of the Negev,
Beer-Sheva 84105, Israel}
\date{\today}

\begin{abstract}
We study the effect of driving a two-dimensional honeycomb system out of equilibrium using
an elliptically polarized light as a time-dependent perturbation. In particular, we try to
understand the topological phase diagram of this driven system when the external drive is a
vector potential given by ${\bf A}(t) = (A_{0x} \cos(\Om t), A_{0y} \cos(\Om t + \phi_0))$. 
These topological phases are characterized by the Floquet Chern number which, in each of
these phases, is related to the number of robust edge modes on a nanoribbon. We show that
varying the ratio $A_{0x}/A_{0y}$ of the external drive is a possible way to take the system
from a trivial to a topological phase and vice versa.
\end{abstract}

\maketitle
\section{Introduction}

One of the most promising playgrounds for both experimental and theoretical studies of
topological phases in two-dimensional systems is the wonder material graphene and other related
materials with similar honeycomb lattice structure. While pristine graphene is a gapless and
trivial system, taking into account the effects of spin-orbit coupling (SOC) in graphene, can
generate topologically non-trivial phases. Experimentally, such an SOC can be induced by
proximity to a topological insulator (TI) such as bismuth selenide \cite{Kou2013,zhang2014} or
by functionalizing graphene using methyl \cite{zoll2016}. As in all topological systems
\cite{hasan, BHZ1,Moore1, Moore2, fu1}, such phases are characterized by and insulating bulk,
hosting conducting boundary modes. A topological invariant (for instance, a Chern number in the
case of two-dimensional TIs) is derived from the bulk bands, and defines the properties of the
boundary modes via the bulk boundary correspondence.

While the physics of topological phases and edge states in graphene is by itself quite
intriguing to the generation of such phases by driving the system out of equilibrium
\cite{kit2010, kit2011, oka2009, gu2011, lin2011, su2012,kun2014,dora2012,thak2013,
kat2013, zhu_2014,rud2013,lin2015, car2015, xiong2016, thak2017, muk2018, zhou2018,
lopez12, zhang16} is a rapidly developing field of research. Using an external
perturbation periodic in time - either by varying a parameter periodically with time or
by using polarized light - it is  possible to modify the behavior of the system and
change its topological character. In particular, we can generate topological phases 
by applying a time-periodic drive to a system which was non-topological to begin with. 
Equally interesting is the possibility of manipulating or destroying the topologically
non-trivial nature of a system by using an external drive with appropriate parameters.
Since we assume that these out-of-equilibrium systems are perfectly periodic in nature,
we employ Floquet theory \cite{floq1883} to study their behavior.

There are several studies that use circularly polarized light \cite{rjs2} to generate and/or
modify Floquet topological phases. However the more general case of elliptical polarization
remains largely unexplored to the best of our knowledge. Even though the result of using an
elliptically polarized light \cite{seshell22,kit21,bayuk21,Chnafa2021,zhu_2014,fern2020} might
appear qualitatively somewhat similar to the effect of using circular polarized light, there
are some major differences. The main reason for these differences is the anisotropy which the
elliptical polarization introduces into the system. 

The plan of this paper is as follows. We begin in Sec. \ref{sec:Equil} with an overview of the spectrum
and edge modes of a honeycomb lattice in presence of a staggered potential. This is followed in Sec.
\ref{sec:Ell} by a brief description of elliptically-polarized light which is then used in Sec.
\ref{sec:Floq} to drive the honeycomb lattice . We find that Floquet topological phases depend not only
on the frequency and amplitude of the drive  but also on the relative phase between the components of the
field. These phases are identified using the Chern number which is calculated using the Floquet eigenstates.
Finally, in Sec, \ref{sec:edge} we demonstrate the bulk boundary correspondence for
these phases by studying the Floquet edge modes that lie on the boundary of a nanoribbon fashioned out
of a honeycomb lattice system. The number of these edge states is found to be consistent with the Chern
number of the bulk system.

\section{Pristine graphene at equilibrium} \label{sec:Equil}
Spinless electrons on a honeycomb lattice are governed by the two-band momentum-space tight-binding
hamiltonian 
\beq
H = \sum_{\vk} \begin{pmatrix}a_{\vk}^\dag&b_{\vk,\da}^\dag\end{pmatrix} \non
h(\vk)
\begin{pmatrix}a_{\vk} \\ ~\\b_{\vk} \end{pmatrix},
\eeq
where $a_\vk$ and $b_\vk$ denote the annihilation operators for $A$ and $B$
sublattice respectively and,
\bea 
h(\vk) = \begin{pmatrix} +\mu_S &~& f(\vk) \\ \\ f^*(\vk) & ~ &-\mu_S \end{pmatrix} \label{eq:h2}.
\eea

\begin{figure}[H]
\sfig[]{\ig [height=7cm]{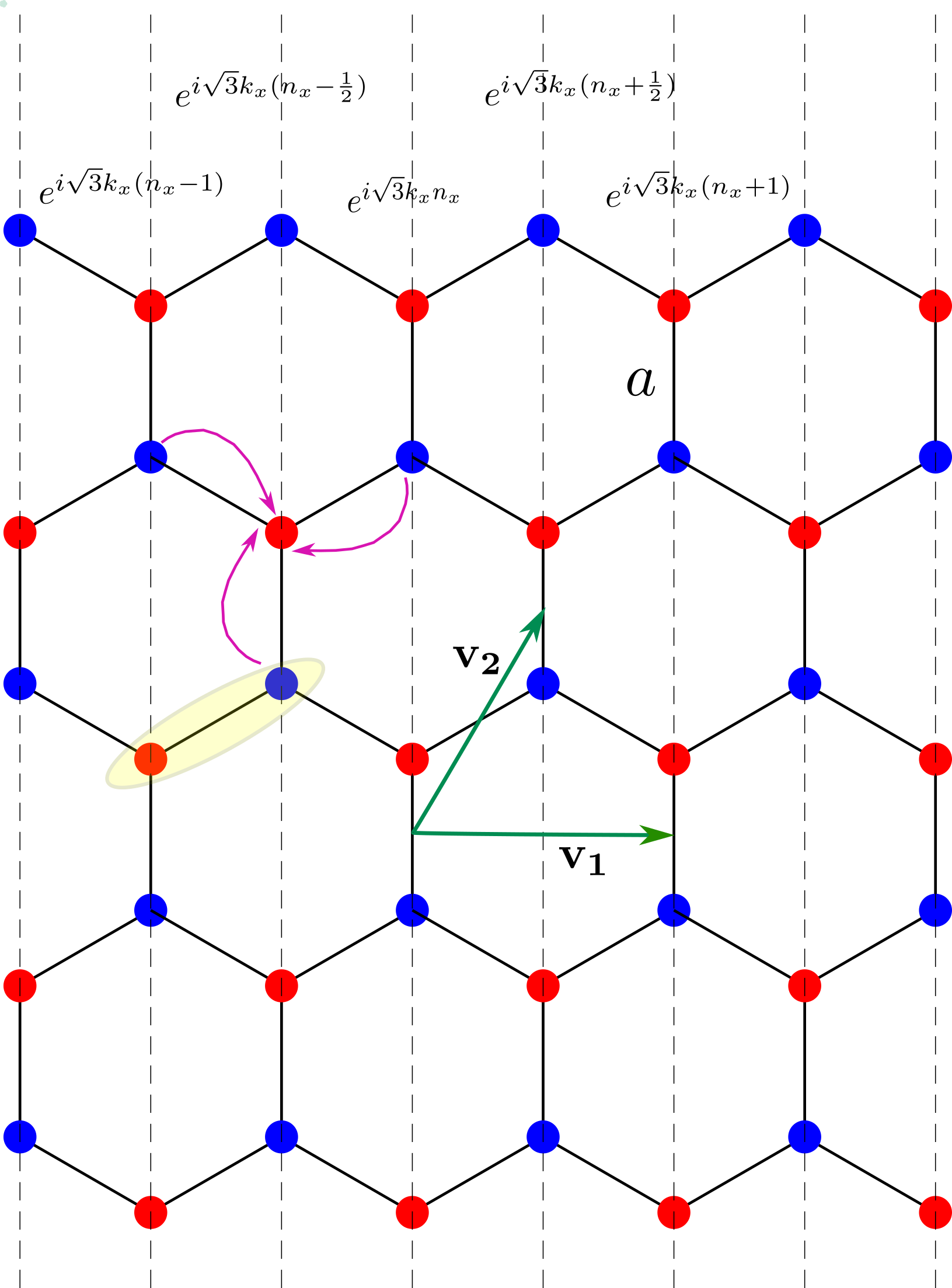}\label{fig:hexlattice}}
\sfig[]{\ig [height=7cm]{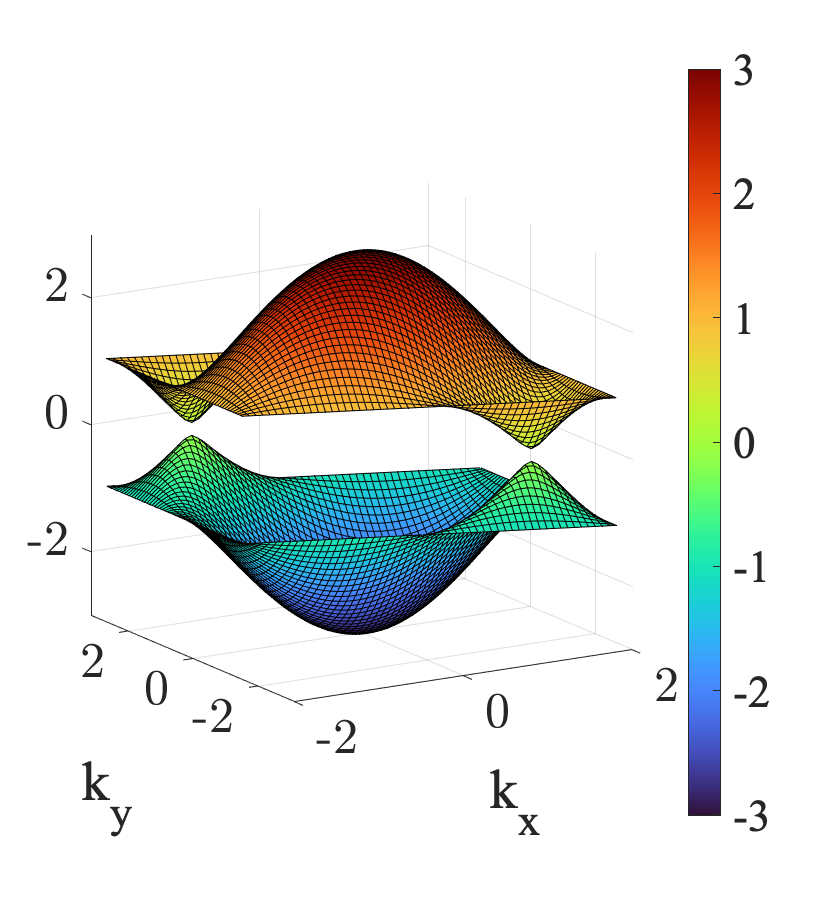}\label{fig:EqSpec}}
\sfig[]{\ig [height=7cm]{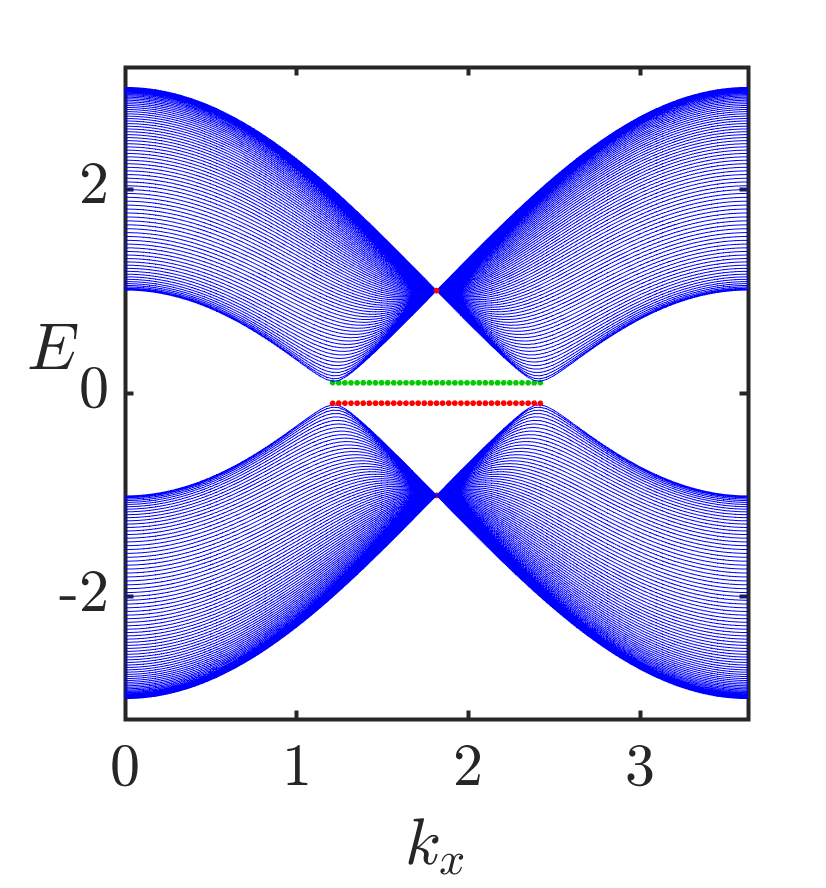}\label{fig:EqRibbon}}
\caption{(a) A honeycomb lattice is a triangular lattice with a two-site unit cell of type $A$ (red)
and $B$ (blue). The spanning vectors of the underlying triangular lattice are given by green arrows
$\bf{v_1}$ and $\bf{v_2}$. The pink arrows show the nearest neighbor hoppings. Note that the zigzag
edge is parallel to the $x-$direction. The bottom edge of the ribbon has sites of type $A$ whereas
the top edge has sites of type $B$. (b) The bulk spectrum in presence of a staggered potential $\mu_s$,
has a band gap $\delta E = 2\mu_S$ at the $K$ and $K'$ points. (c) The edge state spectrum shows that
there are edge modes (red at top edge and green at bottom edge) which are not robust.}
\end{figure}

The off-diagonal terms denote nearest neighbor hoppings with strength $\ga$,
\bea
f(\vk) = \ga\Big(1+e^{i \vk\cdot\bf{v_2}}+e^{i \vk\cdot\bf{(v_2-v_1)}}\Big) = 
\ga \Big(1 + 2\cos\frac{\sqrt{3}k_x}{2} e^{3ik_y/2}\Big).
\eea
The vectors ${\bf v_1}$ and ${\bf v_2}$ are the spanning vectors of the lattice in real space as
shown in Fig.\ref{fig:hexlattice}, and $\mu_S$ is a staggered potential, also referred to as the
Semenoff mass \cite{bern2006,Sem1984}. The lattice parameter $a$ ($\approx 0.14$ nm for graphene)
is set to unity. {\color{black} All the energy scales are measured in units of the nearest-neighbor
hopping strength $\ga$.}

{\color{black}
Diagonalizing this Hamiltonian,
\beq
h(\vk) |\psi_\pm(\vk)\ra = E_\pm(\vk)|\psi_\pm(\vk)\ra 
\eeq
we get the two-band spectrum shown in Fig. \ref{fig:EqSpec}. When the staggered
potential $\mu_S$ is absent, the spectrum is gapless with two Dirac points in the Brillouin zone
(B.Z.) at $K(K') = (\pm{2\pi/3\sqrt{3}},\mp{2\pi/3} )$. The two-component spinors $|\psi_\pm(\vk)\ra$
are then used to define the Chern number of the upper (lower) band as
\bea 
C_\pm =~ \frac{i}{2\pi} ~\int \int dk_x dk_y \Big[ ~
\frac{\partial \psi_\pm^\dg}{\partial k_x} \frac{\partial \psi_\pm}{\partial k_y} -
\frac{\partial \psi_\pm^\dg}{\partial k_y} \frac{\partial \psi_\pm}{\partial k_x} \Big].
\label{eq:chern}
\eea

In order to numerically calculate the Chern number we follow the procedure in described in App.
\ref{sec:app_num}.

}

When the staggered potential is absent, i.e. in case of pristine graphene the spectrum is gapless
and the system is topologically trivial. 
Introducing a non-zero $\mu_s$ gaps out the spectrum as shown in Fig. \ref{fig:EqSpec},
and keeps the system topologically trivial, i.e. the Chern numbers $C_{\pm}$ of the
upper (lower) band is zero. {\color{black} According to bulk boundary correspondence, the
topological invariant, in this case the Chern number, determines the number of boundary modes on
a finite sample of the material.}

This {\color{black}topologically trivial nature} is reflected in the boundary modes of a ribbon.
To study this we consider the system in
a strip geometry, i.e. infinite along the $x-$direction which is parallel to the zigzag edge
and finite with $Ny=80$ sites along the $y-$direction. 
The top edge has all sites of the type $B$, whereas the bottom edge has sites
of type $A$. Since there is a translational symmetry along the $x$ direction, we use $k_x$ as
a good quantum number and write an effective one-dimensional Hamiltonian for a
finite chain running along the $y-$ direction, connected with the subsequent chain by
plane-wave factors as shown in Fig. \ref{fig:hexlattice}. This is given by
\bea
h^{1D}(k_x) &=& \mu_S \sum_{ny}\Big[a_{n_y}^\dag a_{n_y}-b_{n_y}^\dag b_{n_y}\Big]
+ \ga \sum_{n_y} \Big[a_{n_y}^\dag b_{n_y-1} + 2\cos{\frac{\sqrt{3}k_x}{2}}a_{n_y}^\dag b_{n_y}\Big].
\label{eq:h1d}
\vspace{-0.4cm}
\eea

This gives us the spectrum shown in Fig. \ref{fig:EqRibbon}. The band formed by the bulk states are
shown in blue while the green and red colors denote states on the bottom and top edge respectively.
As is clear from the figure, these edge states do not cross from one band to the other. Therefore,
any boundary modes that appear along the edge of a nanoribbon are not robust to perturbations and
can be done away with by introducing a small disorder in the system. 

Note that this staggered potential is different from a Kane-Mele mass \cite{KM2005} which, in
addition to introducing a band gap, transforms the system to a $Z_2$ topological insulator
\cite{sesh16,sesh17} and gives rise to robust edge states.

\section{Elliptically polarized light}\label{sec:Ell}
The most general form of the time-dependent vector potential of elliptically polarized
light is given by
\bea 
\bf{A}(t) &=& (A_{0x}\cos(\Om t),A_{0y}\cos(\Om t + \phi_0)). \label{eq:ell_vec} 
\eea
Here $\phi_0$ denotes the phase difference between the $x$ and $y$ components of the
time-dependent fields. The electric field is therefore written as,
\bea
{\bf E}(t) = -\frac{\partial {\bf A} }{\partial t} = 
(E_{0x} \sin(\Om t),E_{0y}\sin(\Om t + \phi_0)), \label{eq:ell_Efld}
\eea

where $E_{0x(y)} = \Om A_{0x(y)}$. We note that both linear and circular
polarization are special cases of Eq.~\eqref{eq:ell_Efld}. 

\begin{figure}[H]
\centering
\sfig[$\phi_0 = \pi/3$]{\ig[width=5cm]{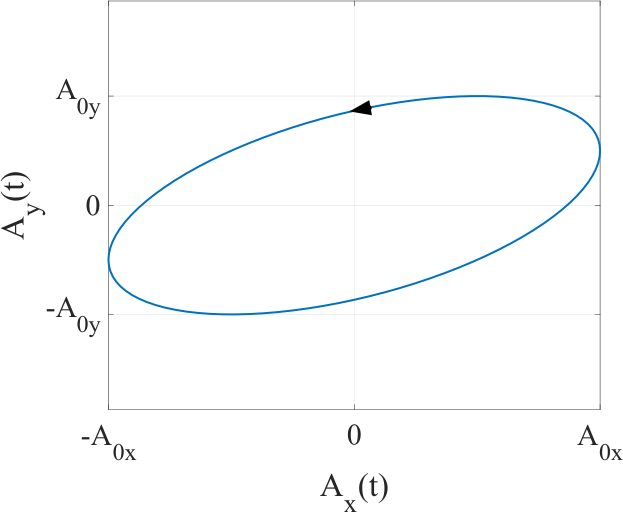}}\hspace{1cm}
\sfig[$\phi_0 = \pi/2$]{\ig[width=5cm]{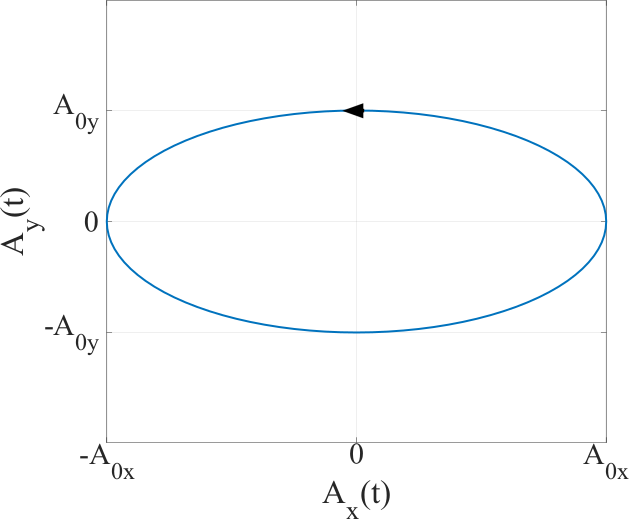}\label{fig:pi2}}
\caption{The polarization ellipse for two values of the phase $\phi_0$ (a)$\pi/3$ and (b)$\pi/2$.
Clearly, when $\phi_0 = \pm \pi/2$, the axes of the ellipse are parallel to the $x$ and $y-$directions.
The special case of $A_{0x} = A_{0y}$ with $\phi_0 = \pm \pi/2$ gives left(right) circular polarization.
For other values of $\phi_0$ the ellipse is rotated. In both these figures we have taken $A_{0x} = 0.7$
and $A_{0y}=0.4$. The ratio of these amplitudes decides the ``flatness" of the ellipse. The ellipticity
or eccentricity of the ellipse depends both on the ratio of amplitudes and the phase.}
\label{fig:ell} \end{figure}

{\color{black} Figure \ref{fig:ell} shows the polarization ellipses for two choices of parameters.}
Fixing $\phi_0=\pm \pi/2$, we obtain elliptically polarized light with the axes
of the ellipse aligned with the cardinal axes as shown in Fig. \ref{fig:pi2}. Further, if 
$A_{0x} = A_{0y}$ and $\phi_0=\pm \pi/2$, we obtain left/right circularly polarized light. 
While the ratio $A_{0x}/A_{0y}$ decides the ``flatness" of the ellipse, the ellipticity
depends on this ratio as well as the phase.

The vector potential given by Eq.~\eqref{eq:ell_vec} enters the momentum-space Hamiltonian
via minimal coupling, $\vk \longrightarrow \vk+{\bf A}$. Therefore, the bulk Hamiltonian 
in Eq.~\eqref{eq:h2} is modified as $h(\vk) \longrightarrow 
h(\vk+{\bf A})$.

\section{Floquet topological phases}\label{sec:Floq}
\begin{figure}[htb]
\begin{center}
\sfig[]{\ig[width=6cm]{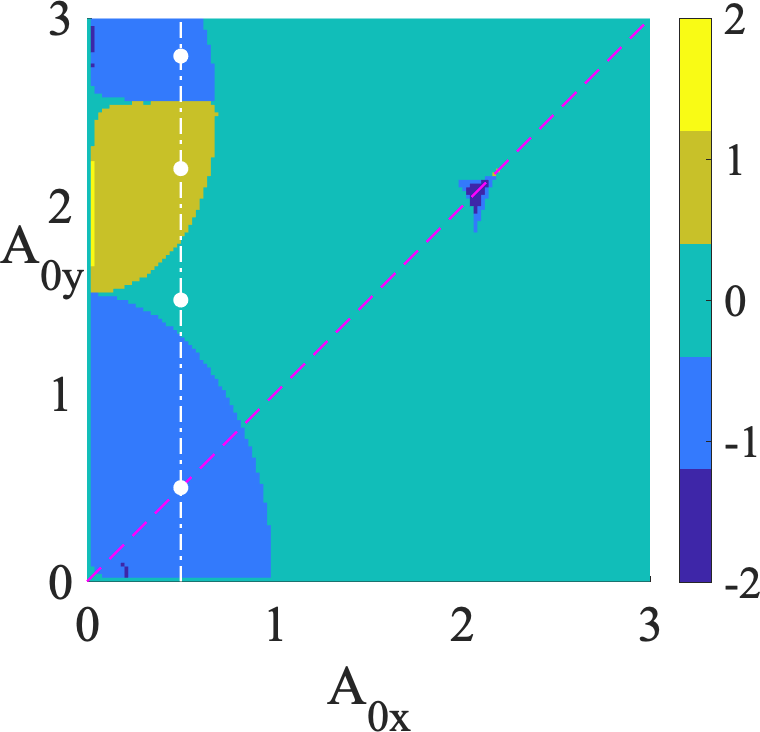}\label{fig:piby2}}\hspace{1cm}
\sfig[]{\ig[width=6cm]{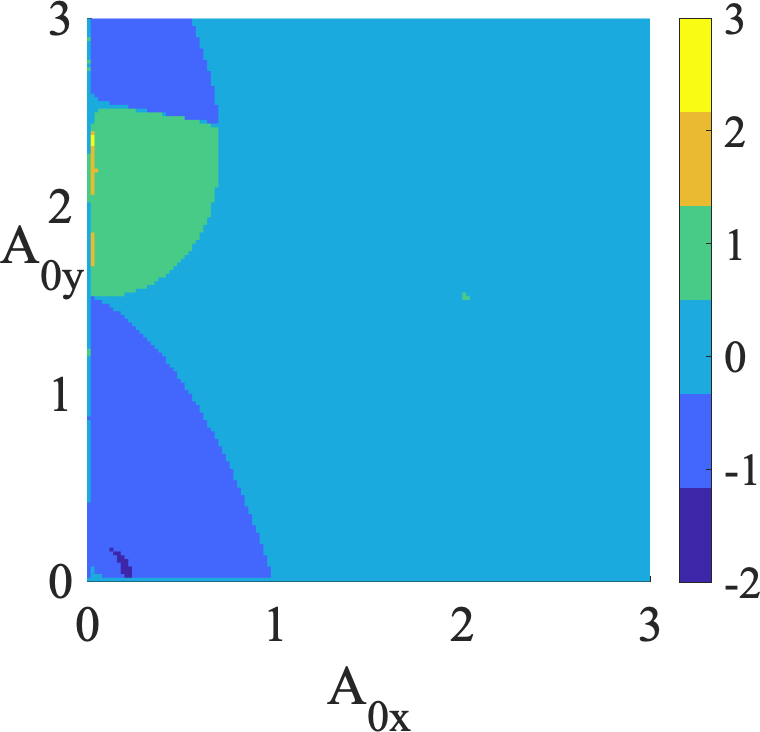}\label{fig:piby3}}
\end{center}
\vspace{-0.5cm}
\caption{Phase diagram showing the Chern numbers $C_+$ of the top band as a function of drive
amplitudes for two different values of the phase (a) $\phi=\pi/2$ and (b)$\phi=\pi/3$. We take
the staggered potential $\mu_s = 0.001$ i.e., we start from a gapped but trivial phase and
drive using a frequency $\Omega = 3$ which is much larger than the band gap of the spectrum
of the unperturbed system. The pink dashed line $A_{0x} = A_{0y}$ in (a) is the
special case of circularly polarized light. The vertical white dot-dash line goes through four
different phases by keeping $A_{0x}$ constant and varying $A_{0y}$. The four dots on this line
correspond to the parameters chosen to study the edge state spectrum in Fig. \ref{fig:ribbon}.}
\label{fig:floq_phases}
\end{figure}

Now we use the form of the perturbation described in Sec. \ref{sec:Ell} to drive the
honeycomb lattice out of equilibrium. We employ Floquet theory to study this since the
drive is assumed to be perfectly periodic in time. Consider a time-dependent Hamiltonian
$H(t)$ with periodicity $T = 2\pi/\Omega$ i.e.,
\beq H(t) = H(t+T), \eeq 
where $\Om$ is the frequency associated with the driven system.
According to Floquet theorem \cite{floq1883, holt}, the solutions to the time-dependent
Schr\"odinger equation (setting $\hbar=1$)
\beq \Big( H(t)-i \frac{\partial}{\partial t} \Big) \Psi(t)= 0, \label{eq:TdepSch} \eeq
are given by
\beq \psi_\al(t) = e^{-i \ep_\al t} \phi_\al (t), \label{eq:ftheory} \eeq
where the quasienergy $\ep_\al$ is unique modulo $n\Om$, i.e.
\beq
\ep_\al \equiv \ep_\al + n\Om, ~~~~ n=0,\pm1\pm2 ...
\eeq
{\color{black} This means that now there are multiple copies of each band. The two bands
with $n=0$ lie in the range $-\pi<\ep_\al T<\pi$ and are called the ``primary Floquet bands",
while the copies of these bands lying outside this energy range are the ``side bands". Note
that this is analogous to the concept of Brillouin Zone in Bloch theory where the periodicity
on a real space lattice gives rise to a Bloch momentum; similarly in Floquet theory the
periodicity in time is reflected in the quasienergy spectrum.}
The state $\phi_\al(t)$ is  periodic with the same time period as the
Hamiltonian $H(t)$, i.e,
\beq \phi_\al(t) = \phi_\al(t+T). \eeq
The time evolution operator from an earlier time $t_1$ to a later time $t_2$ is
defined as
\beq \mc{U}(t_2,t_1) = \mfr{T} e^{-i \int_{t_1}^{t_2} dt H(t)},\non \eeq
{\color{black} where $\mfr{T}$ denotes time ordering and is essential, since the
hamiltonians at two different times do not, in general, commute with each other.}
In particular, for exactly one drive cycle, this time-evolution operator is
called the Floquet operator, i.e.,
\beq \mc{U}_T = \mc{U}(T,0) = \mfr{T} e^{-i \int_{0}^T dt H(t)}.
\label{eq:FlOp} \eeq
Since $\psi(t+T)= \mc{U}_T \psi(t)$, from Eq.~\eqref{eq:ftheory},
\beq \mc{U}_T \psi_\al = e^{-i \ep_\al T} \psi_\al. \eeq
{\color{black}}
We can then use the Floquet eigenstates $\psi_\al$s thus obtained to find the Chern numbers
according to the prescription in App. \ref{sec:app_num} {\color{black} for various drive
parameters. We fix the drive frequency at $\Om = 3$ in units of the hopping parameter
$\gamma$, which is much larger than the band gap of the unperturbed system.
We then plot the Chern number of the upper band (in the primary Floquet zone)
as a function of the drive amplitudes $A_{0x}$ and $A_{0y}$ for various values of the 
phase $\phi$. This gives rise to the phase diagrams of Fig. \ref{fig:floq_phases}.}
An interesting feature of these phase diagrams is the absence of reflection symmetry about the
$A_{0x} = A_{0y}$ line for any value of $\phi_0$. This is strikingly different from the case
of the driven Bernevig-Hughes-Zhang (BHZ) model \cite{seshell22} where the corresponding Floquet
topological phase diagram has the aforementioned reflection symmetry. The reason for this is that
the BHZ model is constructed from a square lattice while the honeycomb lattice has a hexagonal
structure for which the Hamiltonian does not transform trivially under $k_x \leftrightarrow k_y$. 

\section{Edge modes} \label{sec:edge}

Next we verify the bulk boundary correspondence for the phases in Fig. \ref{fig:floq_phases}.
For this, we again consider a nanoribbon which is infinite along the $x-$direction i.e. along the
zigzag edge as shown in Fig.\ref{fig:hexlattice}. Since now we are studying the effect of a polarized
light, we use minimal coupling and Peierls substitution to modify the momentum and the hopping 
parameter respectively as, 
\bea 
k_x &\rightarrow& k_x + A_{0x}\cos(\Om t), \non \\
\ga &\rightarrow& \ga ~ e^{i \bf{A(t)}.\bf{r}},
\eea

where $\bf{r}$ is the vector joining the two sites between which hopping is being considered.
This modifies the Hamiltonian for a nanoribbon as
\bea
h^{1D}(k_x,t) &=& \ga \sum_{n_y} \Big[a_{n_y}^\dag b_{n_y-1} e^{-iA_y(t)}
+ 2\cos{\frac{\sqrt{3}(k_x+A_x(t))}{2}}a_{n_y}^\dag b_{n_y}e^{i\frac{A_y(t)}{2}}\Big].\label{eq:h1dfl}
\eea

\begin{figure}[htb]
\begin{center}
\sfig[$A_{0x} = 0.5,A_{0y} = 0.5$]{\ig[width=4.7cm]{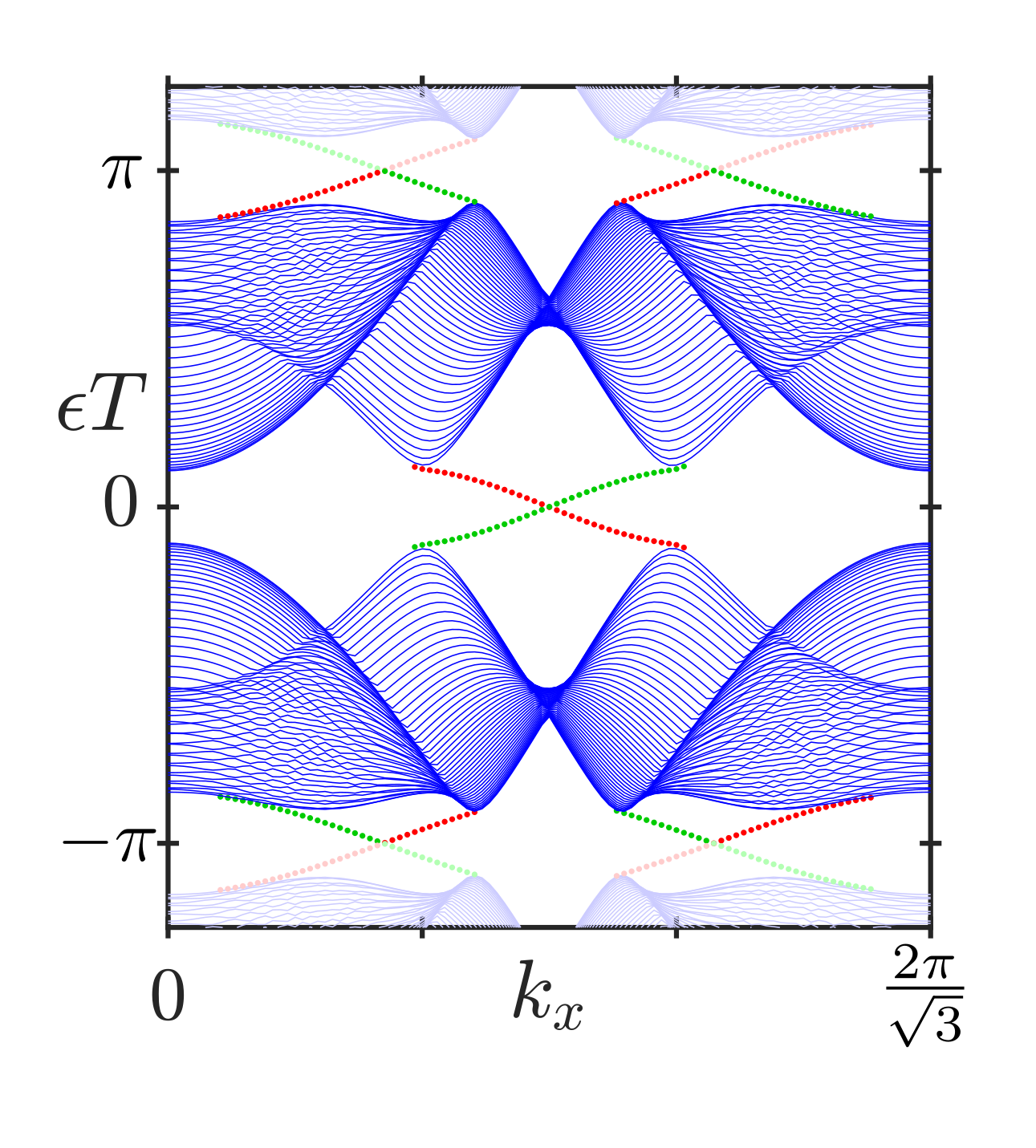}\label{fig:p1}}\hspace{-0.45cm}
\sfig[$A_{0x} = 0.5,A_{0y} = 1.6$]{\ig[width=4.7cm]{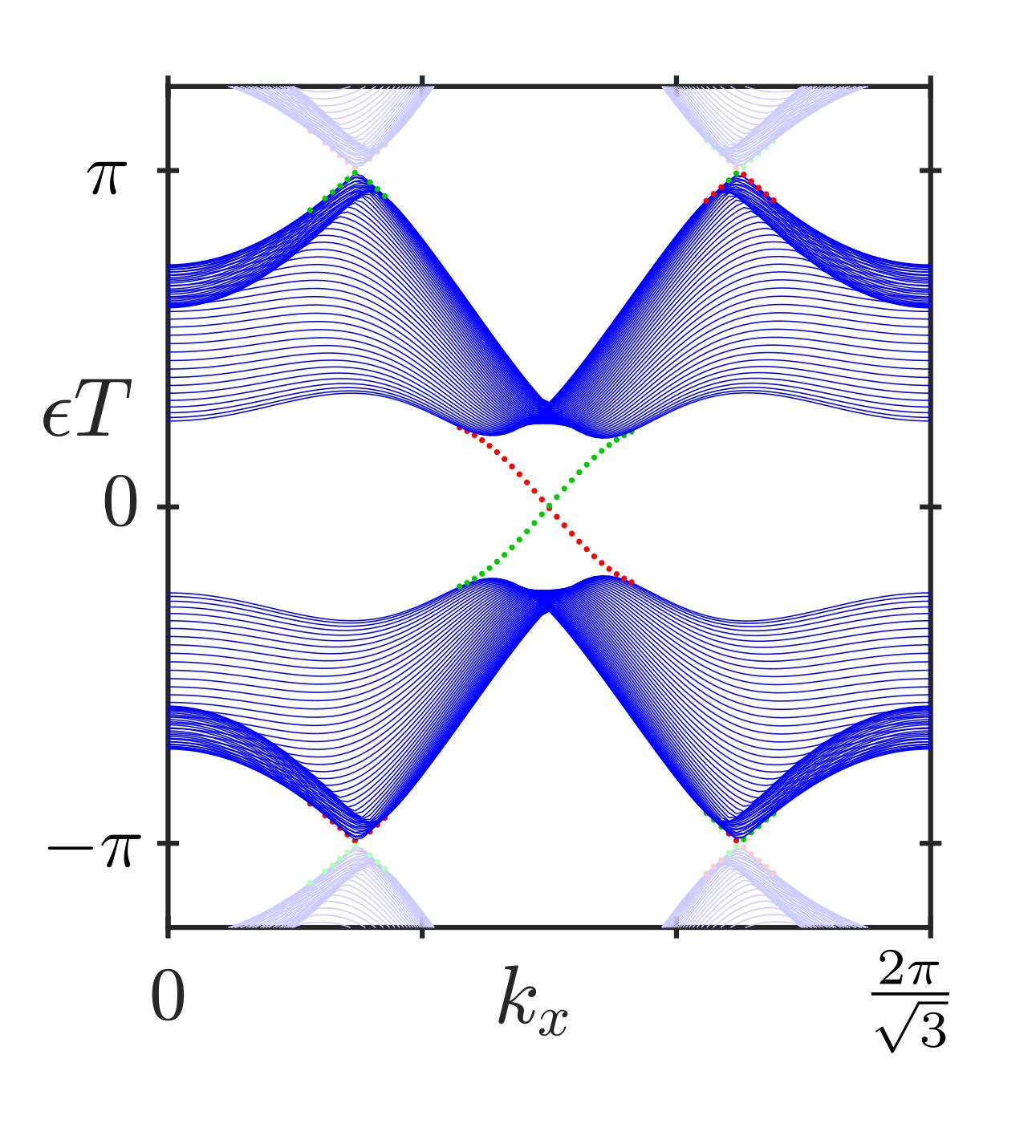}\label{fig:p2}}\hspace{-0.45cm}
\sfig[$A_{0x} = 0.5,A_{0y} = 2.2$]{\ig[width=4.7cm]{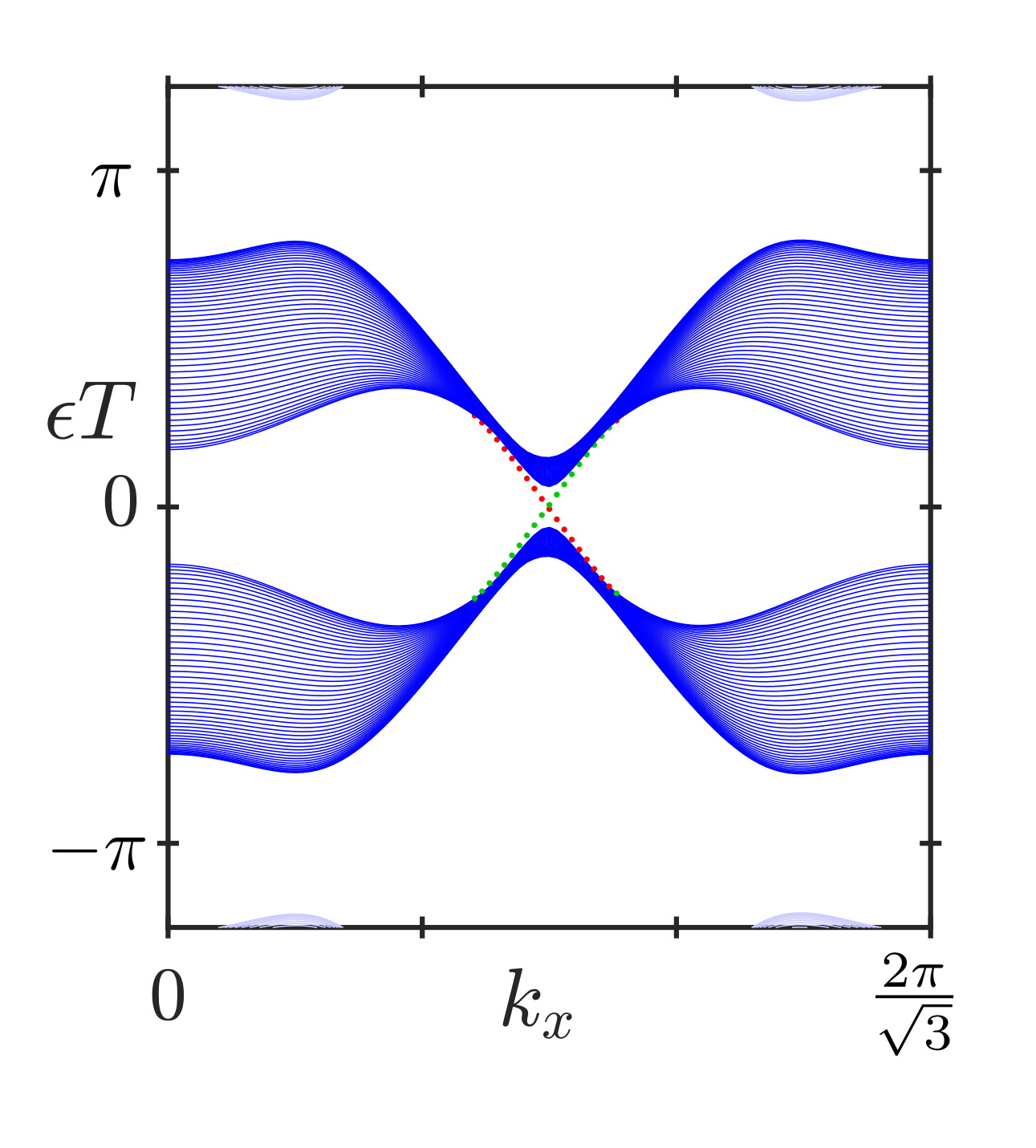}\label{fig:p3}}\hspace{-0.45cm}
\sfig[$A_{0x} = 0.5,A_{0y} = 2.6$]{\ig[width=4.7cm]{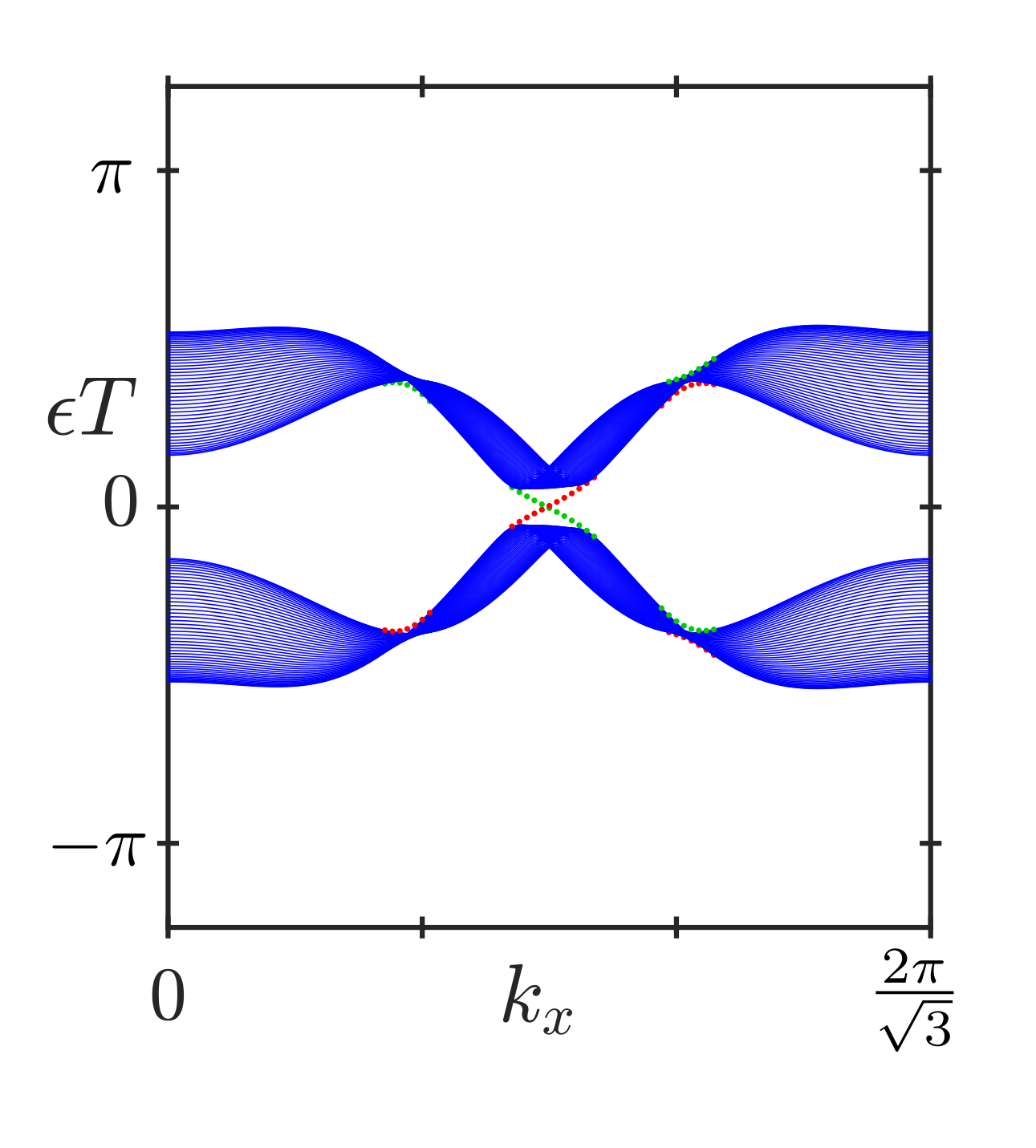}\label{fig:p4}}
\end{center}
\caption{Edge state spectra for different ratios of the drive amplitudes while keeping the
frequency $\Om = 3$ and phase $\phi_0 = \pi/2$ fixed. These correspond to the four points
marked along the white line in Fig. 2(a). The red and green colors correspond to states localized
at the bottom and top edge respectively.}
\label{fig:ribbon}
\end{figure}

We then use the above Hamiltonian to construct the time evolution operator for this one-dimensional
chain, diagonalizing which gives us Floquet eigenvalues and eigenstates. The quasienergy spectrum
thus obtained is shown in Fig. \ref{fig:ribbon}.

We find that the number of edge states are consistent with the Floquet Chern numbers
in the phase diagram of Fig. \ref{fig:floq_phases}. To see this we fix $\phi_0=\pi/2$
and $A_{0x} = 0.5$ and choose values of $A_{0y}$ in the four different phases of
Fig.\ref{fig:piby2}. 
The magnitude of the Chern  number counts the number of edge modes while the sign indicates
the chirality, i.e. the direction of propagation. In Fig. \ref{fig:p1}, we find one pair of
edge states  at $\epsilon T = 0$ and two pairs at $\epsilon T = \pm \pi$. These have opposite
directions of propagation i.e. at $\epsilon T = 0$ we have a top edge right mover while at
$\epsilon T = \pm \pi$ we have two top edge left movers. These add up to give a Chern
number $C_+ = -1$. In Fig. \ref{fig:p2} we seem to have edge states. However there is no
gap at $\epsilon T = \pm \pi$ which makes the system trivial. In Fig.\ref{fig:p3} we find
a top edge right mover corresponding to a Chern number $C_+ = +1$, whereas in Fig.\ref{fig:p4}
we have a top edge left mover corresponding to a Chern number $C_+ = -1$.

\section{Conclusions and Outlook}
We have studied the Floquet topological phases generated by driving a honeycomb lattice such
as graphene out of equilibrium by using polarized light. However, we consider the most general
case of elliptically polarized light. Some of the effects of such a time-dependent perturbation
are markedly different from the more commonly studied cases of linear or circular polarization.
In particular, a rich topological phase diagram is generated with Chern numbers depending
on both the phase $\phi_0$ and the ratio $A_{0x}/A_{0y}$. Keeping $\phi_0$ fixed and varying this
ratio allows us to tune in and out of topological phases. We also find that the chiral edge
modes on a nanoribbon of this driven system are consistent with the Floquet Chern numbers
obtained from the bulk. 

While this system with its rich phase diagram is by itself quite interesting, another aspect to
consider is the effect of spin-orbit coupling on such a system. For instance, it is well known
that a Kane-Mele type SOC makes the equilibrium honeycomb lattice topological. The interplay of
this SOC with an external perturbation using elliptically polarized light could lead to an even
richer phase diagram and could allow us to engineer topological phases with higher Chern numbers.

{\color{black} This work is restricted to studying the topological properties of the Floquet bands
that arise as a result of periodically driving the electrons on a honeycomb lattice. However, due
to the fact that there are infinitely many copies of these bands corresponding to different integer
values of $n$, the occupation of these bands is very different from a conventional Fermi distribution.
The effect of this distribution in the out-of-equilibrium system is expected to play a pivotal
role in observable properties of the material such as the longitudinal and transverse conductance
which are directly related to the topological nature of the system.
}

\appendix
\section{Numerical evaluation of Chern number}\label{sec:app_num}
{\color{black}
Given a hamiltonian $h(\vk)$, the eigenvalue equation is given by
\beq
h(\vk) |\psi_\al(\vk)\ra = E_\al(\vk)|\psi_\al(\vk)\ra 
\eeq
with $|\psi_\al(\vk)\ra$ being the eigenstate corresponding to eigenvalue $E_\al$. In the model
we have considered $\al = \pm$ corresponds to the upper/lower bands and $|\psi_\al(\vk)\ra$s
are two-component spinors. However, note that the discussion that follows holds good for any
$n-$band system (in which case each eigenstate is an $n-$component spinor).

The eigenstates $|\psi_\al(\vk)\ra$ are then used to define the Berry curvature,
and the Chern number of the $\al$th band as
\bea 
C_\al =~ \frac{i}{2\pi} ~\int \int dk_x dk_y \Big[ ~
\frac{\partial \psi_\al^\dg}{\partial k_x} \frac{\partial \psi_\al}{\partial k_y} -
\frac{\partial \psi_\al^\dg}{\partial k_y} \frac{\partial \psi_\al}{\partial k_x} \Big].
\label{eq:chernapp}
\eea
When we numerically evaluate $|\psi_\al\ra$ there is an arbitrary phase factor which is included
thereby making the direct numerical computation of he above expression complicated. We need a method
to calculate this which is gauge-invariant, i.e. independent of the arbitrary phase factor which can
always multiply the eigenvector. Therefore we resort to the method similar to Fukui et. al.
\cite{fukui2005}. 

Consider an actual numerical computation where the eigenvectors are evaluated on a discretized
two-dimensional Brillouin Zone (B.Z). We denote the points on the momentum mesh as $(n_x,n_y)$ with
$n_x\in [1,N_x], n_y \in [1,N_y]$. The $\al$th eigenvector at each such momentum point is written as
$|\psi_{\al,n_x,n_y}\ra$. Four consecutive $k-$points on this momentum mesh form a plaquet.
We define a variable $F_{\al,n_x,n_y}$ associated with this plaquet as

\bea
F_{\al,n_x,n_y}&& \non \\
=&&\ln \Bigg[\frac{\la\psi_{\al,n_x,n_y}|\psi_{\al,n_x+1,n_y}\ra}{|\la\psi_{\al,n_x,n_y}|\psi_{\al,n_x+1,n_y}\ra|} \frac{\la\psi_{\al,n_x+1,n_y}|\psi_{\al,n_x+1,n_y+1}\ra}{|\la\psi_{\al,n_x+1,n_y}|\psi_{\al,n_x+1,n_y+1}\ra|}
\frac{\la\psi_{\al,n_x+1,n_y+1} |\psi_{\al,n_x,n_y+1}\ra}{|\la\psi_{\al,n_x+1,n_y+1} |\psi_{\al,n_x,n_y+1}\ra|}
\frac{\la\psi_{\al,n_x,n_y+1}|\psi_{\al,n_x,n_y}\ra}{|\la\psi_{\al,n_x,n_y+1}|\psi_{\al,n_x,n_y}\ra|}\Bigg]\label{eq:Fplaq}.\non \\
\eea
Note that we use only the principal branch of the logarithm function in Eq. \ref{eq:Fplaq}. 
Summing this over the entire B.Z. gives the Chern number of the $\al$th band, which is an integer
\beq
C_{\al} = \frac{1}{2 i \pi}\sum_{n_x,n_y} F_{\al,n_x,n_y}.
\eeq
}
\bibliography{refs}
\end{document}